\documentclass[aps,pre,showpacs,twocolumn]{revtex4}
\usepackage{graphicx}

\newcommand{\tw}{t_{\scriptstyle \rm w}}
\newcommand{\kB}{k_{\scriptscriptstyle \rm B}}

\newcommand{\tk}{t_{\scriptscriptstyle \rm K}}

\newcommand{\Eeq}{e_{\scriptscriptstyle \rm eq}}

\newcommand{\teq}{t_{\scriptscriptstyle \rm eq}}
\newcommand{\veq}{v_{\scriptscriptstyle \rm eq}}
\newcommand{\eis}{e_{\scriptscriptstyle \rm IS}}

\begin{document}


\title{Kovacs effect in facilitated spin models of
strong and fragile glasses} 

\author{Jeferson J. Arenzon} 
\email{arenzon@if.ufrgs.br}
\affiliation{Instituto de F{\'\i}sica, Universidade Federal do
Rio Grande do Sul \\ CP 15051, 91501-970 Porto Alegre RS, Brazil}
\altaffiliation[Research Associate at ]{The Abdus Salam ICTP, Trieste,
  Italy}

\author{Mauro Sellitto}
\email{sellitto@ictp.trieste.it}
\affiliation{The Abdus Salam International Centre for Theoretical
  Physics \\  Strada Costiera 11, 34100 Trieste, Italy}


\begin{abstract}
  We investigate the Kovacs (or crossover) effect in facilitated
  $f$-spin models of glassy dynamics.  Although the Kovacs hump shows
  a behavior qualitatively similar for all cases we have examined
  (irrespective of the facilitation parameter $f$ and the spatial
  dimension $d$), we find that the dependence of the Kovacs peak time
  on the temperature of the second quench allows to distinguish among
  different microscopic mechanisms responsible for the glassy
  relaxation (e.g. cooperative vs defect diffusion). We also analyze
  the inherent structure dynamics underlying the Kovacs protocol, and
  find that the class of facilitated spin models with $d>1$ and $f>1$
  shows features resembling those obtained recently in a realistic
  model of fragile glass forming liquid.
\end{abstract}

\maketitle

\section*{Introduction}

There is much interest in understanding how the out of equilibrium
properties of glassy materials can be described within a statistical
mechanics framework.  At low temperatures, the exceedingly long
relaxation time prevents such materials from reaching thermal
equilibrium on the accessible time scales, and makes hard to establish
a connection with the thermodynamic phase behavior~\cite{McKenna}.  In
particular, several questions about the nature of the glass state are
still unanswered: for example, does it represent a genuine
thermodynamic phase of the matter or a purely kinetic phenomenon? How
many ``fictive'' thermodynamic parameters are needed for
characterizing the amorphous state? These and other questions are
relevant not only for glasses but for every system with slow
dynamics~\cite{Leticia03}, such as dense granular materials and
colloids, disordered dielectrics and electron glasses, etc.

Fifty years ago, Kovacs designed an experiment in which the low
temperature behavior of glasses is probed by means of the following
protocol~\cite{Kovacs} (see also~\cite{Struik,Larson}).  A sample is
brought from high to low temperature $T_1$, and left to age for a time
$\tw$ until its volume $v$ attains the value it would have in thermal
equilibrium at a temperature $T_2>T_1$, $v(\tw)=\veq(T_2)$. The sample
is then suddenly heated to $T_2$ and the ensuing volume relaxation is
observed.  Even though the average volume is already at its
equilibrium value, it turns out that in order to fully equilibrate,
the system first increases its volume, then passes through a maximum
that depends on the thermal history of the system, and eventually
recovers the equilibrium.  The non-monotonic behavior generally
results from the superposition of two antagonist processes: the fast
modes - equilibrated at $T_1$, tend to increase the volume, while the
slow modes - which are still relaxing towards $T_1$, tend to decrease
it~\cite{Larson}.  The sample is therefore formed by regions having an
average local density larger or smaller than the equilibrium one,
which respond on different time scales to a change in temperature.
The experiment thus provides an indirect evidence that the low
temperature dynamics of glasses is spatially
heterogeneous~\cite{Kovacs,Struik,Larson}.

The Kovacs effect was first observed in polymeric
glasses~\cite{Kovacs} and in statistical mechanical systems as simple
as the ferromagnetic Ising chain~\cite{Brawer}.  There is an
increasing interest in this phenomenon in connection with other kind
of memory effects investigated in disordered magnets, and recent
statistical mechanical approaches to glassy and granular matter.  In
these contexts, it has been observed that the Kovacs effect occurs in
systems with quasi-long range order~\cite{BeHo}, spin
glasses~\cite{BeBo}, realistic model of molecular liquid
(ortho-terphenyl)~\cite{MoSc}, traps models~\cite{Bertin}, kinetically
constrained spin chains~\cite{Boh}, and in simple models of granular
compaction, such as the parking lot model~\cite{TaVi}. More recently,
the effect has also been studied in the spherical $p$-spin
glass~\cite{Leticia}.

In this paper we study the Kovacs effect in facilitated $f$-spin
models of glassy dynamics in spatial dimension $d$.  Our interest here
is mainly dictated by the possibility of using the Kovacs effect to
identify the microscopic mechanism responsible for the slow relaxation
(e.g., {\it strong} vs.  {\it fragile} glass behavior~\cite{Angell}),
or, more ambitiously, to discriminate between situations in which an
arrest transition occurs from others in which it does not.
 We also investigate the inherent structure dynamics during the Kovacs
protocol and show that models in spatial dimension $d>1$ and
facilitation parameter $f>1$ are able to account for a subtle effect
recently observed in molecular dynamics simulation of a realistic
model of fragile glass forming liquid~\cite{MoSc}.

\section*{Facilitated spin models}

The model we consider is 
model first introduced by Fredrickson and Andersen~\cite{FrAn}, with
the purpose of describing the cooperative relaxation in highly viscous
liquids.  The spins are non interacting and the Hamiltonian density is
simply
\begin{equation}
  e = \frac{1}{N} \sum_{i=1}^N s_i \,,
  \label{H}
\end{equation}
where the 
$s_i=0,1$ are spin variables, and the index $i$ runs over the sites of
a lattice with coordination number $c$. One possible interpretation of
the spin variable is that its state represents, after a suitable
coarse-graining over molecular length and time scales, mesoscopic
regions of the liquid with more or less density.  What makes the model
non trivial is the dynamics, which is defined by
the following facilitation rule: $s_i$
can flip with a non-zero rate if and only if at least $f$ of its $c$
neighboring spins are up, that is
\begin{equation}
  h_i \equiv \sum_{k \, {\rm nn} \, i} s_k  \geq f\,. 
\label{gamma}
\end{equation}
The transition rates satisfy detailed balance and are proportional to
the Metropolis factor
\begin{equation}
    {\cal W}(s_i\to 1-s_i)  =  \Theta \left( h_i-f \right)
    \min \left[ 1, \, {\rm e}^{ \beta (2s_i-1)}  \right] \,,
    \label{W_s}
\end{equation}
where $\beta$ is the inverse temperature $\beta =1/\kB T$, and the
Heaviside function $\Theta$ encodes the facilitation mechanism.  It
turns out that at low enough temperature the system dynamics is unable
to relax to the equilibrium distribution within the simulation time
window.  In fact, as the density of up spins becomes smaller and
smaller, the facilitation rule is hardly satisfied and consequently
the spin relaxation becomes slower and slower. For this reason the
model exhibits several glassy features, such as non-exponential
relaxation, history dependent properties and heterogeneous
distribution of local relaxation times~\cite{FrAn,FrBr,BuHa,Grant}.
For a review see Refs.~\cite{RiSo03}.  This type of facilitated spin
model is reminiscent of the kinetically constrained lattice-gas
introduced by Kob and Andersen~\cite{KoAn}. However, the facilitated
dynamics is not conserved and the constraint is weaker (as it involves
only the local field $h_i$ of the spin which attempts to
flip). Numerical investigations of two and three dimensional systems
(with different values of $f>1$) suggested that the characteristic
equilibration time at low temperature diverges as a Vogel-Fulcher
law~\cite{Grant}, although early arguments shown that the system is
ergodic at any temperature larger than zero~\cite{FrBr,Reiter}.  More
generally, the equilibration time has a non-Arrhenius behavior typical
of fragile glasses for $d>1$ and $f>1$.  The case $f=1$ instead
corresponds to the defect diffusion dynamics and leads to the
Arrhenius behavior, $\log \teq \sim 1/T$, characterizing strong
glasses.  However, $1d$ models, such as the East
model~\cite{East-model,SoEv99} which have directed (or asymmetric)
constraints, may also exhibit a non-Arrhenius behavior.

One reason of interest in the facilitated spin models is that they
present many glassy features in spite of the absence of a true glass
transition in finite dimension~\cite{Reiter,ToBiFi}, while on the
Bethe lattice they appear to have an ergodic/non-ergodic transitions
of the kind predicted by mode-coupling
theory~\cite{ReMaJa,PiYoAn,SeBiTo}.  In the following we shall be
interested in reproducing the Kovacs experiment in facilitated spin
models defined on finite dimensional lattices with the purpose
of comparing the behavior for different value of the facilitation
parameter.

\subsection*{Kovacs effect}

To numerically reproduce the Kovacs experiment an equilibrium
configuration at high temperature is quenched at low temperature $T_1$
and left to age for a waiting time $\tw$, until the energy density,
$e$, attains the value it would have in thermal equilibrium at
temperature $T_2>T_1$, $e(\tw) = \Eeq(T_2)$. According to 
Eq.~(\ref{H}) the equilibrium energy density reads:
\begin{equation}
    \Eeq(T)  =  \frac{1}{1+{\rm e}^{\beta}}      \,.
    \label{E_eq}
\end{equation}
The energy density therefore is the natural observable that plays in
our model the role of the volume in the Kovacs protocol.  This
quenching procedure is repeated for several temperatures $T_1$. Each
configuration obtained at time $\tw$ such that $e(\tw) = \Eeq(T_2)$,
represents a distinct glassy state with different thermal history but
identical energy. Such configurations are then quickly heated at
temperature $T_2 > T_1$, and the reduced energy $\Delta v$, defined as
\begin{eqnarray}
    \Delta v(t,\tw) & = & \frac{e(t) - e(\tw)}{e(\tw)} \,,
    \label{Delta_v}
\end{eqnarray}
is recorded as a function of time $t$ ($t> \tw$). We use a fast
algorithm that keeps track of the list of {\it free} (unconstrained)
spins, from which the attempted spin-flips are chosen with the usual
Metropolis algorithm. Accordingly, the Monte Carlo time sweep is
rescaled with the number of mobile spins.  The largest number of
spins we consider is $N=2^{21}$.  The lattice systems have periodic
boundary conditions.

\begin{figure}
\begin{center}
\includegraphics[width=8.5cm]{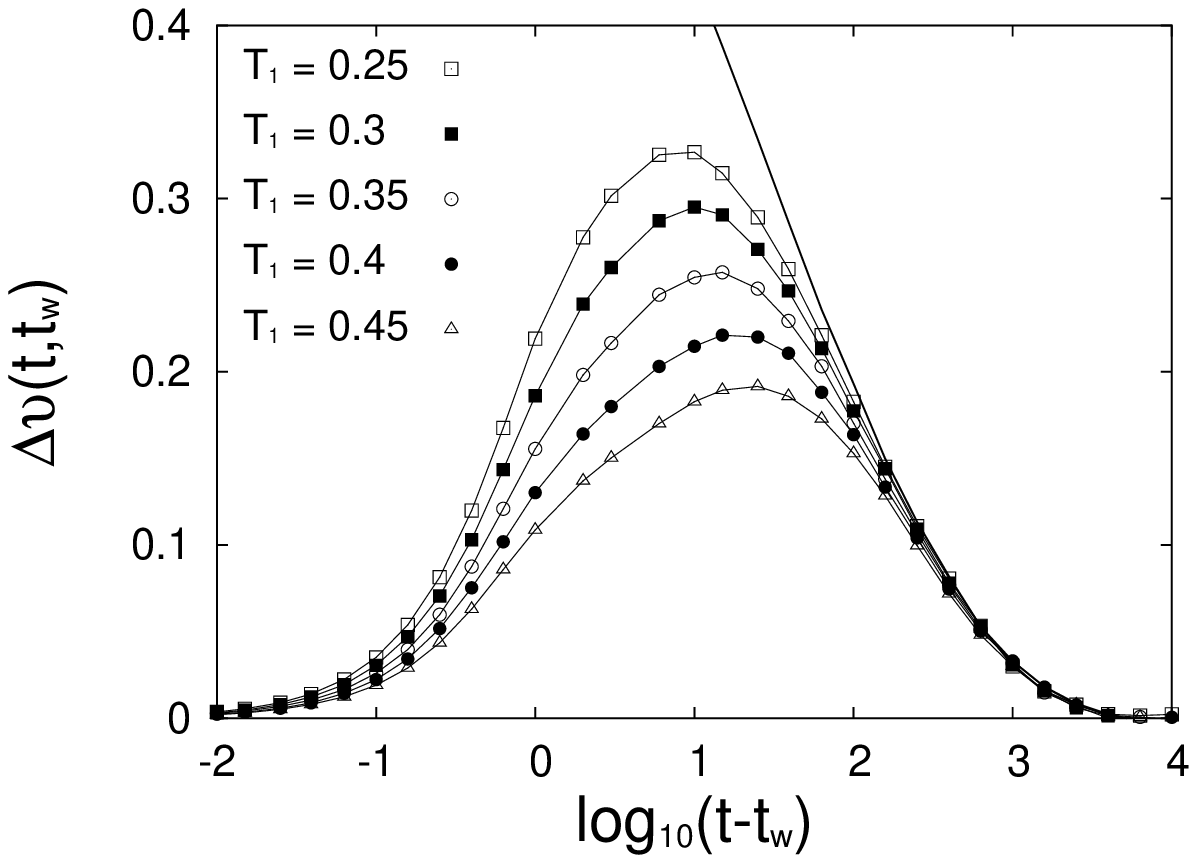}
\includegraphics[width=8.5cm]{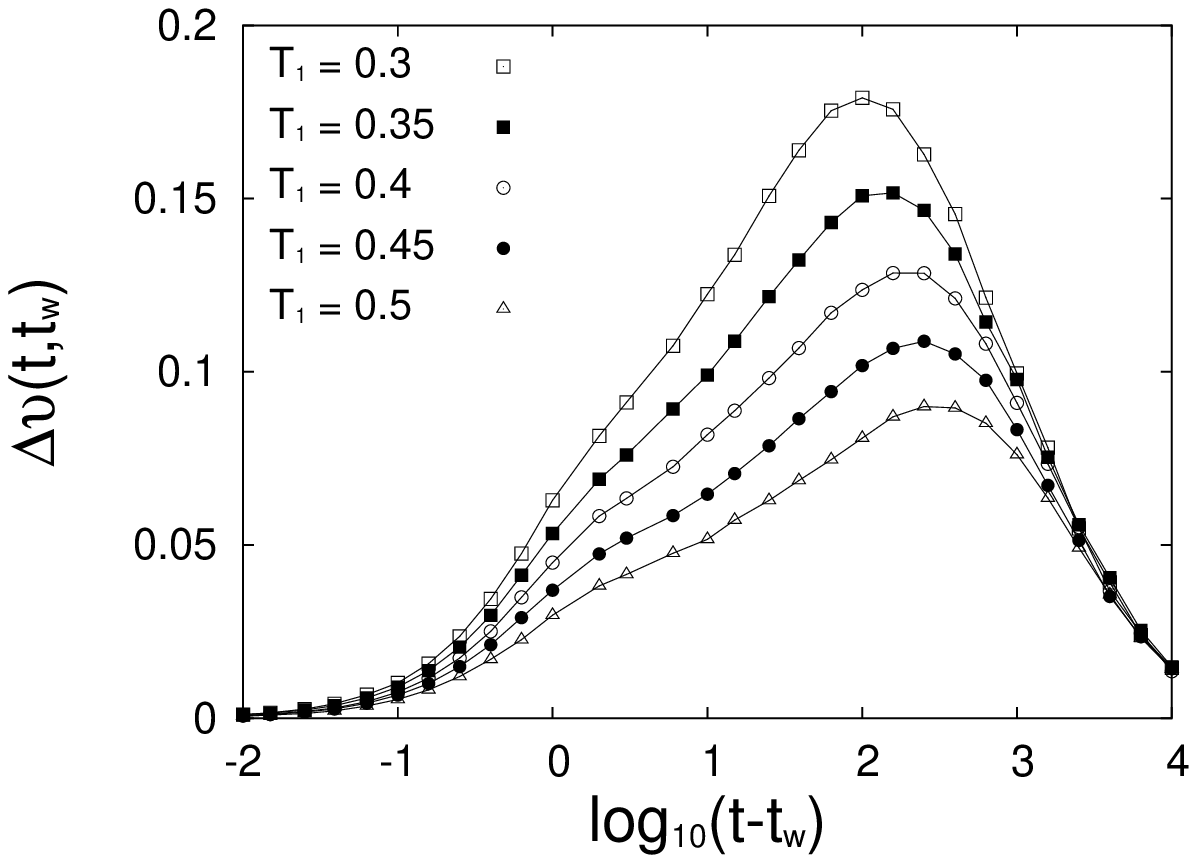}
\end{center}
\caption{Kovacs effect in a facilitated spin model on a cubic lattice
  with facilitation parameter $f=3$ (system size $V=128^3$). The
  system relaxes at temperature $T_1$ until it attains the energy
  $e(\tw) = \Eeq(T_2)$; afterwards the temperature is changed from
  $T_1$ to $T_2=1.0$ (top) or 0.8 (bottom).  
  The full line represents the energy relaxation
  after a quench at $T_2$.
  Qualitatively similar results are found for different
dimensions, facilitations and geometries.}
\label{kov_3d}
\end{figure}

\begin{figure}
\begin{center}
\includegraphics[width=8.5cm]{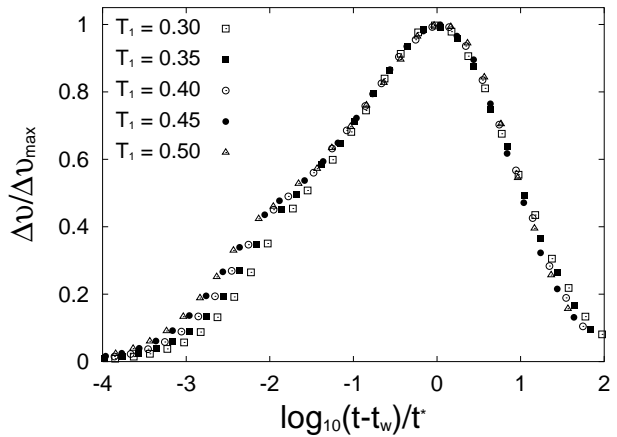}
\end{center}
\caption{Scaling of the data of Fig.~\ref{kov_3d} for
$T_2=0.8$. Both time and height are rescaled, respectively
by $t^*$ (see text) and $\Delta v_{\scriptstyle\rm max}$. Data collapse 
is good except for
the initial part where scaling is not expected to hold~\cite{BeBo}.}
\label{scaling}
\end{figure}

In Fig.~\ref{kov_3d} (top) we show the reduced energy relaxation
during a numerical experiment for a $3d$ facilitated spin system with
$f=3$, final temperature $T_2 = 1.0$ and several initial quenching
temperatures $T_1$.  Qualitatively similar results are obtained for $d=3$ 
and smaller facilitation parameters ($f=1,\,2$), for the square lattice with
$f=1,\,2$, as well as on a Bethe lattice (a random graph with fixed
connectivity) where the glass transition occurs at a finite
temperature for $f>1$~\cite{SeBiTo}.
The plot exhibits the typical Kovacs hump 
very similar to those obtained in many other 
systems~\cite{Brawer,BeHo,BeBo,MoSc,Bertin,TaVi,Boh,Leticia}, the hump height
increasing with $\Delta T=T_2-T_1$ and shifting to smaller times as
$T_1$ decreases. 
By decreasing $\Delta T$ one observes (see Fig.~\ref{kov_3d}, bottom,
where $T_2=0.8$) the clear two-step behavior of the energy relaxation,
as recently found in the $1d$ case~\cite{Boh}.  The fast, non
activated processes due to the excitations of spins neighboring
facilitated ones, contribute to the formation of a plateau starting on
timescales~\footnote{In $1d$, this corresponds to the nucleation of
smaller clusters inside the previously formed domains, defined as a
sequence of contiguous down spins, plus the left up
spin~\cite{SoEv99}.  A simple mean field argument taking this into
account, is able to correctly describe both the linear behavior at
short times and the height of the plateau.}  $t-\tw\sim {\cal O}(1)$.
On longer timescales, due to activated processes, the energy increases
up to a maximum at time $t-\tw= \tk$, decreasing afterwards until the
equilibrium is eventually attained.  When rescaling time by a suitable
characteristic time, $t/t^*$, and $\Delta v$ by the maximum height,
$\Delta v_{\scriptstyle\rm max}$, a good collapse of the late time
regime is obtained, while deviations occur for the non activated part
of the hump, as can be seen in Fig.~\ref{scaling}. Usually $t^*$ has
to be chosen much larger than the time of the maximum, $\tk$, so that
the contribution of the fast, non scaling modes
vanishes~\cite{BeBo}. Here, instead, as a consequence of the clear
separation of time scales, $t^*$ is very close to $\tk$.
%

\begin{figure}
\begin{center}
\includegraphics[width=8.5cm]{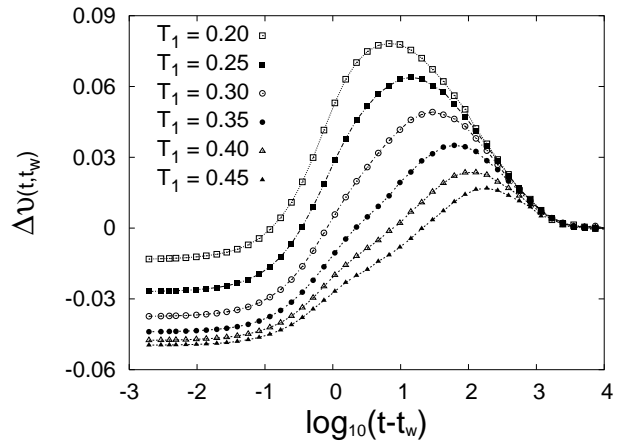}
\end{center}
\caption{Kovacs effect in a facilitated spin model on a cubic
  lattice with $f=3$ (system size $V=128^3$). The protocol here is
  different from fig.~\ref{kov_3d}: the system relaxes at temperature
  $T_1$ for a time $\tw=1000$ and then the temperature is changed
  from $T_1$ to $T_2=1$. See also Ref.~\cite{MoSc}.} 
\label{scior}
\end{figure}

Another possible way to investigate the Kovacs effect is to fix
the time spent at temperature $T_1$ and then to change the temperature
to $T_2$.  An example is shown in Fig.~\ref{scior} for $d=f=3$.  This
protocol has been used in Ref.~\cite{MoSc}.  One can notice that the
results are qualitatively similar to those obtained with the previous
protocol.

Finally, we must remark that, in order to remove the trivial part of
the Kovacs effect, one should consider the initial time slightly after
the jump as discussed in Ref.~\cite{Bertin}. However, for the class of
models considered here the contribution of fast modes, lasting for
${\cal O}(1)$ MCs, is clearly resolved due the presence of an
intermediate plateau.  Moreover, this initial increase can be removed
by looking at the inherent structures dynamics.  We also checked that
within our protocol there is no Kovacs effect in the absence of
kinetic constraints, confirming that the observed effects are not an
artifact of our procedure.

\subsection*{Temperature dependence of the Kovacs peak}

The results we have have obtained for different values of the
facilitation parameter $f$, would lead to the conclusion that the
Kovacs effect is unable to discriminate among the several microscopic
mechanisms of slow relaxation or, even worse, between situations in
which a dynamical arrest transition occurs from others in which it
does not. As we show below, this is not the case: a more careful
analyzes shows clear differences between all these cases.  In
particular, we have analyzed the behavior of the Kovacs peak time
$\tk$ as a function of the final temperature $T_2$ at fixed $T_1$,
making a comparison with recent estimations of the equilibration times
coming from independent methods.
 
For $d=1$ there are simple arguments~\cite{Boh} that correctly
describe the dependence of both $\tw$ and $\tk$ on $T_1$ and $T_2$.
The average distance between up spins is proportional to the inverse
of their concentration, $e^{-1}(T)\sim \exp(1/T)$, for low $T$, while
the effective diffusion rate for these spins is $\Gamma(T)\sim
\exp(-1/T)$. Thus, after the initial quench to $T_1$, the time $\tw$
to attain the equilibrium concentration (at $T_2$) of up spins, $\Eeq
(T_2)\sim \exp(-1/T_2)$, is $\tw(T_1,T_2)\sim \Gamma^{-1}(T_1)
\Eeq^{-2} (T_2) \sim \exp(1/T_1+2/T_2)$.  When $T_1=T_2$ we obtain the
equilibration time after a quench, $\teq\sim \exp(3/T_2)$. A similar
argument, for the East model, leads to $\teq\sim \exp(1/T_2^2\log 2)$
and $\tw\sim \exp(1/T_1T_2\log 2)$. In both cases, it turns out that,
as a function of $T_2$, $\tk\sim\teq$. However, it is not clear
whether simple arguments exist for other dimensions and facilitation
parameters, although the correspondence between $\tk$ and $\teq$ seems
to remain valid, as is shown below.



%
%
Fig.~\ref{fig.scaling_tk} (top) shows $\tk$ vs $T_2$ for the defect
diffusion dynamics, $f=1$, and spatial dimension $d\leq 4$.  In all
cases, $\tk$ shows an Arrhenius behavior in the limit $T_2\to T_1^+$
and scales with $T_2$ as the equilibration relaxation time of the
system. For $d=1$ we recover the result of Ref.~\cite{Boh}, with $\tk
\sim \exp(3/T_2)$. For higher dimensions, a generalization of the
previous argument~\cite{RiSo03} gives $\teq\sim\exp[(1+2/d)/T_2]$, and
seems not to be correct (dashed lines in figure). Indeed, for $d=2$,
the result $\tk \sim\exp(2.3/T_2)$ is compatible with many,
independent numerical estimates of $\teq$, for example, from the
integrated correlation function and the dynamical
susceptibility~\cite{Se03} and from the distribution of persistence
times~\cite{Arenzon04}.  This value is in the same universality class
of directed percolation (DP)~\cite{Hinrichsen00}. For $d=3$ and 4, a
renormalization group analysis~\cite{WhBeGa03} gives
$\teq\sim\exp[(2+\varepsilon/12)/T_2]$, with $\varepsilon=4-d$, again
in the same universality class of DP.

\begin{figure}[floatfix]
\includegraphics[width=8.5cm]{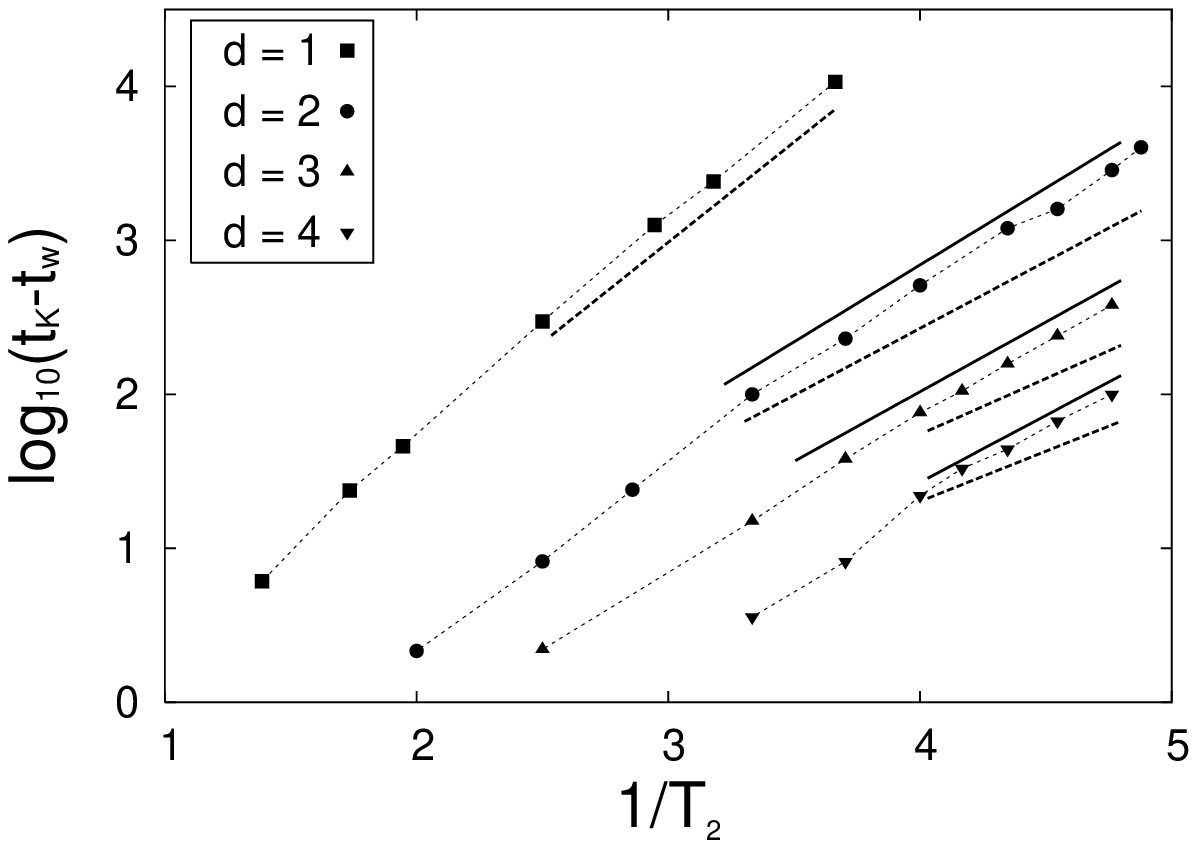}
\includegraphics[width=8.5cm]{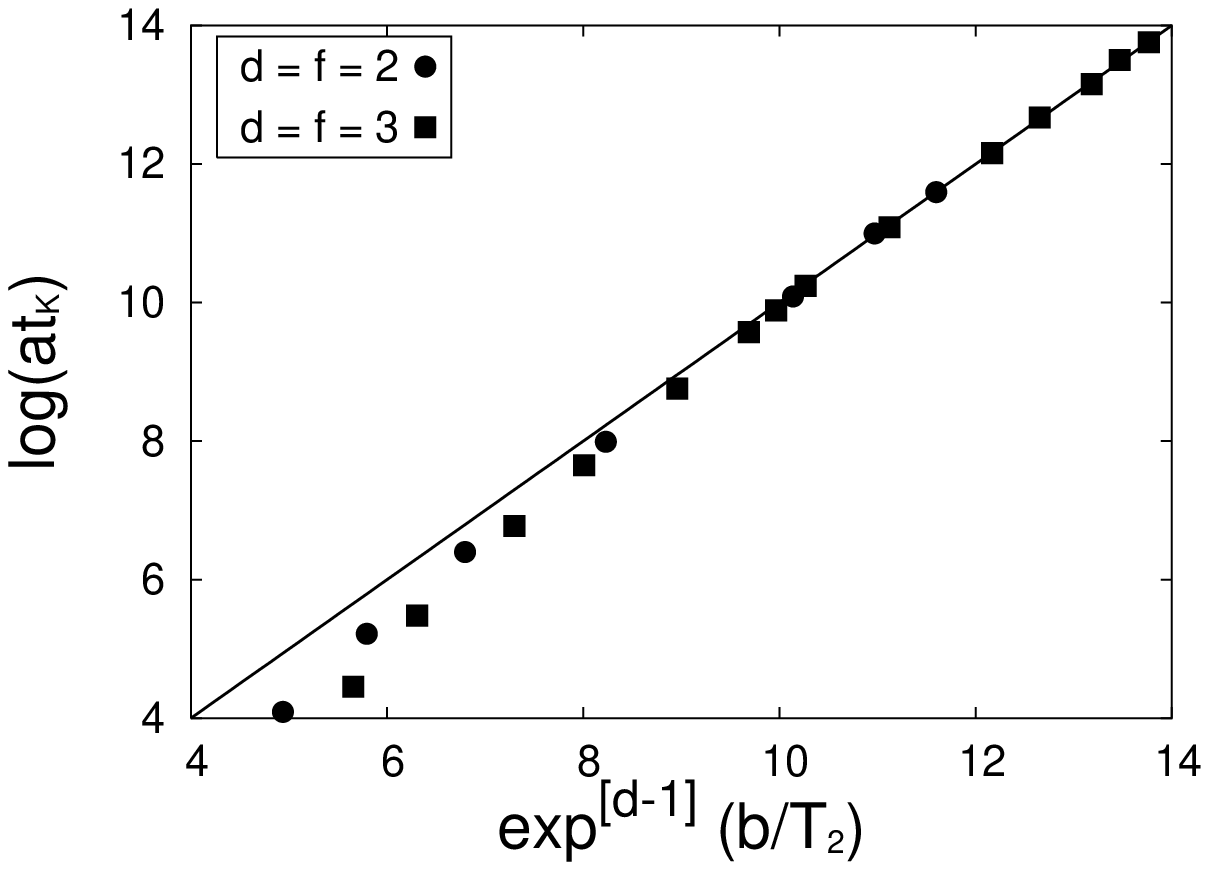}
\caption{Behavior of $t_{\scriptscriptstyle K}$ as a function of
  $T_2^{-1}$.  Top: $f=1$ and $T_1=0.2$.  In all cases the
  behavior, for $T_2\to T_1^+$ is Arrhenius. For comparison we show
  several predictions available in the literature. The solid lines
  correspond to: 
  $\tk\sim\exp(2.3/T_2)$~\cite{Se03,Garrahan,Arenzon04} ($d=2$) and
  $\tk\sim\exp[(2+\varepsilon/12)/T_2]$~\cite{WhBeGa03} with
  $\varepsilon=4-d$ ($d=3$ and 4). The dashed lines correspond to the mean
  field result~\cite{Boh,RiSo03}: $\tk\sim\exp[(1+2/d)/T_2]$.  Bottom:
  $d=f>1$, showing super Arrhenius behavior (see text). Notice
  the iterated exponential in the abscissa: for example, $\exp^{[2]}(x)
  \equiv \exp(\exp(x))$.}       
\label{fig.scaling_tk}
\end{figure}


In Fig.~\ref{fig.scaling_tk} (bottom) we show the results for the
cooperative diffusion dynamics with $d=f>1$. In this case the systems
exhibit a fragile glass behavior. Accordingly, we checked the Kovacs
peak against an iterated exponential function $\tk \sim
\exp(\exp(b_2/T))$ for $d=f=2$, and $\tk \sim \exp(\exp(\exp(b_3/T)))$
for $d=f=3$~\cite{BuHa,Ja,RiSo03,ToBiFi}. Again, our estimations agree 
pretty well, within the
numerical accuracy, with those of the equilibrium relaxation time,
obtained independently from the integral of the correlation function
and the peak of the dynamical susceptibility~\cite{Se03}. For
the East model, whose dynamics is also cooperative, again there is a
correspondence with the relaxation time of the system,
%
A first conclusion that we can draw from this analysis is that 
the Kovacs protocol is rather sensitive to the strong/fragile
relaxation behavior and seems to provide an alternative, independent 
method to obtain the equilibration time of the system.

Some extra information can also be obtained from $\tw$, the time spent
at $T_1$.  In all cases we have considered here, the hump time, $\tk$,
and $\tw(T_1,T_2)$, are related through a power-law $\tk \sim \tw^b$,
where $b$ is temperature independent if $d=f$, and symmetric
facilitation. On the other hand, for $d>f=1$ and in the East model,
$b\sim 1/T_2$ and $b\sim T_2$, respectively.
In domain growth models one finds: $\tk \sim \tw^{1/2}$ (2d XY~\cite{BeHo}) 
and $\tk \sim \tw^{z/(1+z)}$ for the Ising model~\cite{Bertin}. 
For $d=f=1$ we find
$b\simeq 0.68$, close to the expected $2/3$ because in $1d$ the
dynamics proceeds as a domain growth process with dynamic exponent
$z=2$ (we also obtain $2/3$ from the expressions for $\tk$ and $\tw$
derived in Ref.~\cite{Boh}).  On the other hand, for $d=f=2$ and 3 we
get $b \simeq 0.89$ and 1.05, respectively, close to unity, as in the
$p$-spin-glass model~\cite{Leticia}, where $b \simeq 0.9$.

\begin{figure}[floatfix]
\includegraphics[width=8.5cm]{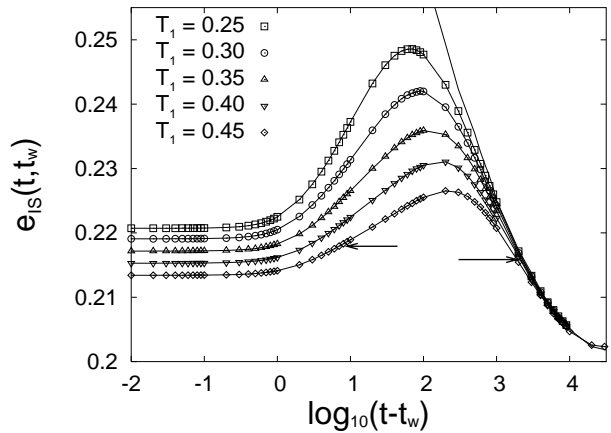}
\caption{Energy density of inherent structures during the Kovacs
  protocol for $d=f=3$ and several values of $T_1$.  The lattice side
  is $L=64$ and $T_2=0.8$.  
  The arrows show two ISs (with energy $\eis=0.2179$ and 0.2158)
  obtained by a zero-temperature descent from configurations having
  identical energy $e = 0.24$ before and after the peak.}
\label{fig.inherent}
\end{figure}

\subsection*{Inherent Structure Dynamics}

Finally, in order to make contact with recent results obtained by
molecular dynamics simulation, Ref.~\cite{MoSc}, we investigated the
inherent structure (IS) dynamics~\cite{StWe83} during the Kovacs
protocol.  For spin models with stochastic dynamics such as those
considered here, an IS is a blocked configuration, i.e. a
configuration in which only isolated defects are present and no
spin-flip is possible.
The IS dynamics is obtained as follows: at
every time $t$ a copy of the system is made and evolved with a
zero-temperature sequential dynamics until it reaches a configuration
in which no further spin-flip is possible.
%
%
The energy $\eis$ of this IS  is
recorded and the dynamics of the original systems is resumed, this
procedure being repeated along the system evolution.
Fig.~\ref{fig.inherent} shows a typical example of IS dynamics for
$d=f=3$.  First, one observes that even if the energies at
time $\tw$ are the same for all $T_1$, 
$e(\tw)=\Eeq(T_2)$, the corresponding inherent structures
are different:
the lower is $T_1$, the higher is $\eis(\tw)$.
%
%
For times up to ${\cal O}(1)$, the IS evolution is confined in the same
``energy basin'' since the fast thermal excitations are removed by
the zero-temperature descent. On longer time scales, the slow, activated
processes start to play a role, the system is able to overcome the
energy barriers and leave the initial basin, given origin to the
hump. After the hump, the curves merge with the one obtained from a 
direct quench to $T_2$ (thick line in Fig.~\ref{fig.inherent})
and eventually reach a value
different (lower) than the initial one, that only
depends on the equilibrium energy, $\eis(\Eeq)$ (e.g., for
$f=1$, $\eis(\Eeq)=\Eeq(1-\Eeq)^d$~\footnote{In $d=1$, as one sequentially
sweeps the lattice, every up spin that has an up spin to its right (those
at left were already flipped down) will become 0. The remaining up spins,
occurring with probability $\Eeq(1-\Eeq)$,
contribute to $\eis$. This is simply generalized in higher dimensions.}).
Thus, in analogy with the
results of ref.~\cite{MoSc}, here too
the IS explored far from equilibrium differ in depth from those
sampled close to equilibrium. This difference 
increases with $T_2-T_1$
and is due to a different number of excitations, or equivalently, to a
different realization of spatial heterogeneities, and not to
topological properties of distinct regions of the phase space. 

\begin{figure}[floatfix]
\includegraphics[width=8.5cm]{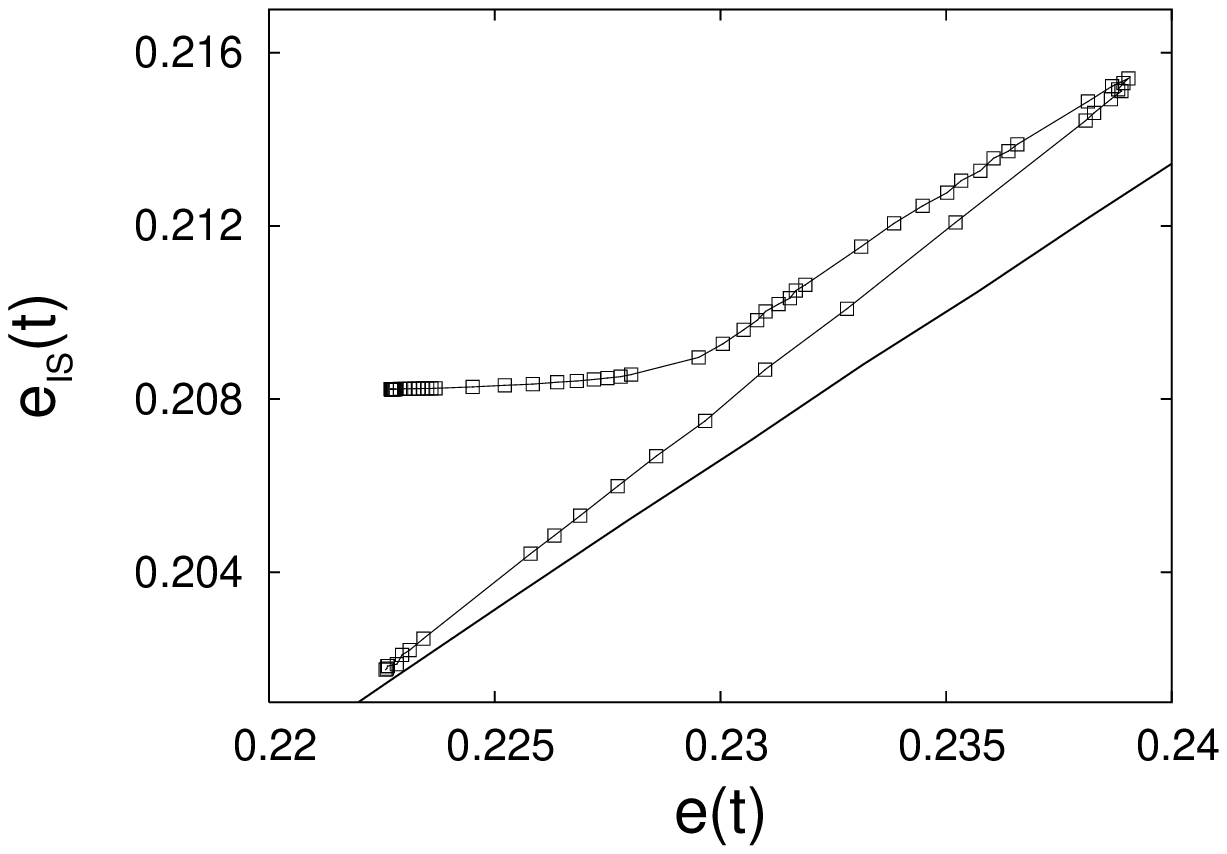}
\includegraphics[width=8.5cm]{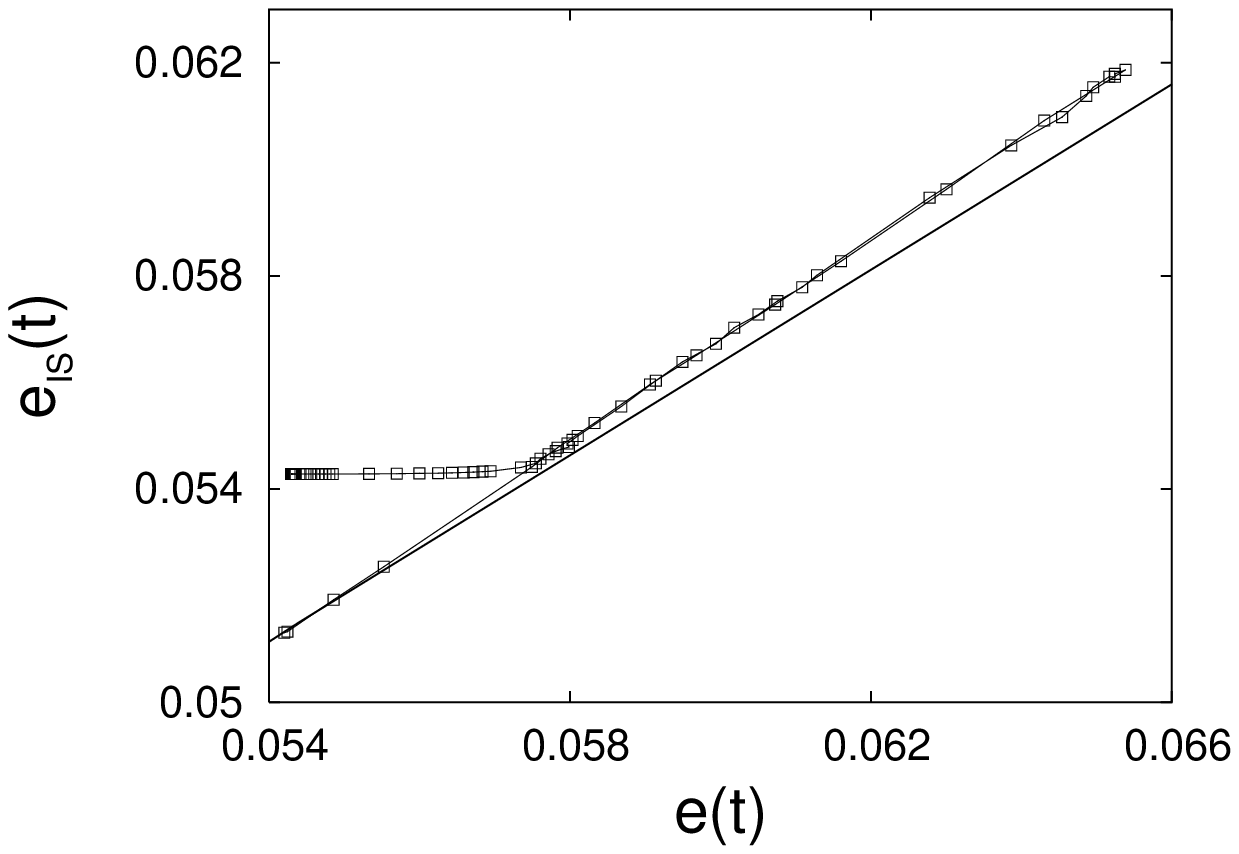}
\caption{Energy density of inherent structures during the Kovacs
  protocol versus the energy density. Top: $d=f=3$ for $T_1=0.6$ and
  $T_2=0.8$. The thick line is the equilibrium value, and both curves
  merge only at the end of the Kovacs protocol.  Bottom: $d=f=1$ for
  $T_1=0.4$ and $T_2=0.7$. Notice that in this case, there is no
  difference between configurations with the same energy on both sides
  of the hump. For late times the curve merges with the equilibrium
  one (thick line), $e_{\scriptscriptstyle \rm IS}^{\scriptscriptstyle
  \rm eq} =\Eeq(1-\Eeq)^d$.}
\label{fig.inherent2}
\end{figure}

Once the system leaves the initial IS, two different behaviors are
possible when one considers a couple of ISs descending from
configurations having the same energy before and after the peak.
While in the simplest case the two ISs may have the same energy there
could be a more complicated case for which this is not true.  The
latter case has been observed recently by Mossa and Sciortino in
molecular dynamics simulation of orto-terphenyl~\cite{MoSc}, a typical
fragile glass forming liquid. We find that for $f>1$ there is an IS
energy-shift similar to that observed in Ref.~\cite{MoSc} (see
Fig.~\ref{fig.inherent2}, top, and the arrows in
Fig.~\ref{fig.inherent}). For 1-spin facilitated dynamics instead, the
ISs have the same energy (see Fig.~\ref{fig.inherent2}, bottom, for an
example).  The reason of this distinct behavior does not seem
directly related to the form of the relaxation time, but rather to the
value of the facilitation parameter $f$. In fact, we find that in the
East model, whose relaxation time is super-Arrhenius, no IS
energy-shift is observed either~\footnote{We also mention that the IS
energy-shift occurs in directed models in higher dimensions, such as
the North-East model.}. Our results seem to suggest that only
facilitated spin models with $d>1$ and $f>1$ are able to reproduce
the subtle effect observed in Ref.~\cite{MoSc}.

In Fig.~\ref{fig.inherent2} we also show the IS sampled in equilibrium
(thick lines) and, interestingly, in all cases, during the whole
protocol the system explores regions never visited while in
equilibrium: only asymptotically the points merges with the
equilibrium curve. However, for smaller temperatures, the separation
between the lower branch and the equilibrium curve diminishes, as well
as the difference between the branches.  Notice that in
Ref.~\cite{MoSc}, soon after the hump, the system starts to sample
configurations from a region of the phase space that is explored in
equilibrium (but at different temperatures).

\section*{Conclusions}

In summary, we have studied the Kovacs effect in a class of spin
models with facilitated dynamics which exhibits slow relaxation at low
temperature.  Our results show that the Kovacs effect is qualitatively
similar in all cases examined, irrespective of the spatial dimension
the facilitation parameter, and the presence or not of a dynamic
transition.  However, in the class of facilitated models considered
here one may distinguish between these situations by looking at
temperature (of the final quench, $T_2$) dependence of the Kovacs peak
time, which scales in the same way as the equilibrium relaxation time
of the system at that temperature.  Thus, the Kovacs protocol provides
an independent way to gain information about the nature of the
relaxation dynamics, at least for this class of models.  It would be
interesting to understand to what extent such a correspondence holds
in more realistic systems, or in systems where the structural
relaxation is related with the topography of the underlying potential
energy landscape.

A related question concerns the inherent structures visited while the
system follows the Kovacs protocol. Generally, they show the non
trivial, activated character of the Kovacs effect, as well as the role
played by spatial heterogeneities.  In contrast with Ref.~\cite{MoSc},
the IS visited by the facilitated models considered here are not the
equilibrium ones, which are only attained asymptotically.  Moreover,
the East and facilitated 1-spin models are not able to reproduce
the subtle IS energy-shift observed in Ref.~\cite{MoSc}, while such an
effect occurs in facilitated spin models with $d>1$ and $f>1$. This
suggests that the inherent structure dynamics underlying the Kovacs
protocol and the IS energy-shift might be used as a tool for
discriminating between distinct microscopic mechanisms of dynamic
facilitation.  It would be also interesting to check by molecular
dynamics simulation whether similar results hold for realistic models
of strong and fragile glasses.  This could provide a more stringent
test of different scenarios of glassy dynamics.

%
%

\section*{Acknowledgments}

JJA acknowledges the ICTP for support and hospitality, the Brazilian
agencies CNPq and FAPERGS for partial support, and R.M.C. de Almeida,
J.P. Garrahan, Y. Levin and D.A. Stariolo for useful conversations.




\end{document}